\documentclass[twocolumn,showpacs,showkey,superscriptaddress,preprintnumbers,amsmath,amssymb,longbibliography]{revtex4-1}
\usepackage[bookmarks=true,colorlinks=true,citecolor=blue,linkcolor=blue,urlcolor=magenta]{hyperref}

\usepackage{amssymb}% http://ctan.org/pkg/amssymb
\usepackage{fdsymbol}

\usepackage{bm}% bold math
\usepackage{xfrac}
\usepackage{calc}
\usepackage{graphicx}
\usepackage[tight,footnotesize]{subfigure}
\usepackage{color}
\usepackage{longtable}%for tables that spans more than one page 
\usepackage{dashrule}%used for dashed lines in tables
\usepackage{dcolumn}% Align table columns on decimal point
\usepackage{multirow}%
\usepackage{rotating}%to turn words in the table cells

\usepackage{float}
\usepackage{perpage}
\usepackage{morefloats}%The package will allocate more “float skeletons” than LaTeX does by default

\usepackage{latexsym}
\usepackage{gensymb} 
\usepackage{bbm}
\usepackage{paralist}

\usepackage[dvipsnames]{xcolor}
\ifpdf
\graphicspath{
	{figures/}
	{figures/pdf/}	
}
\else
\graphicspath{
	{figures/}
	{figures/pdf/}		
}
\fi

\DeclareFontFamily{OT1}{pzc}{}
\DeclareFontShape{OT1}{pzc}{m}{it}{<-> s * [0.900] pzcmi7t}{}
\DeclareMathAlphabet{\mathpzc}{OT1}{pzc}{m}{it}

\newcommand\scalemath[2]{\scalebox{#1}{\mbox{\ensuremath{\displaystyle #2}}}}

\begin{document}
\preprint{AIP/DM}
\title[]{Bounds from multi-messenger astronomy on the Super Heavy Dark Matter}

%\author{M. Deliyergiyev}\email[]{maksym.deliyergiyev@unige.ch}
\author{M. Deliyergiyev}\email[]{maksym.deliyergiyev@outlook.com}
\affiliation{Department of Nuclear and Particle Physics, University of Geneva, CH-1211, Switzerland}

\author{A. Del Popolo}\email[]{adelpopolo@oact.inaf.it}
\affiliation{Dipartimento di Fisica e Astronomia, University Of Catania, Viale Andrea Doria 6, Catania, I-95125, Italy}
%\affiliation{INFN Sezione di Catania, Via S. Sofia 64, Catania, I-95123, Italy}
\affiliation{Institute of Astronomy, Russian Academy of Sciences, 119017, Pyatnitskaya str., 48 , Moscow, Russia}

\author{Morgan~\surname{Le~Delliou}}
\email[Corresponding author: ]{(delliou@lzu.edu.cn,) Morgan.LeDelliou.ift@gmail.com}
\affiliation{Institute of Theoretical Physics, School of Physical Science and Technology, Lanzhou University, No.222, South Tianshui Road, Lanzhou, Gansu 730000, China}
\affiliation{Instituto de Astrof\'isica e Ci\^encias do Espa\c co, Universidade de Lisboa, Faculdade de Ci\^encias, Ed. C8, Campo Grande, 1769-016 Lisboa, Portugal}
\affiliation{Lanzhou Center for Theoretical Physics, Key Laboratory of Theoretical Physics of Gansu Province, Lanzhou University, Lanzhou, Gansu 730000, China}

\date{\today}

\begin{abstract}

The purely gravitational evidence supporting the need for dark matter (DM) particles is compelling and based on Galactic to cosmological scale observations. Thus far, the promising weakly interacting massive particles scenarios have eluded detection, motivating alternative models for DM. We consider the scenarios involving the superheavy dark matter (SHDM) that potentially can be emitted by primordial black holes (PBHs) and can decay or annihilate into ultrahigh-energy (UHE) neutrinos and photons. The observation of a population of photons with energies $E\ge 10^{11}$ GeV would imply the existence of completely new physical phenomena, or shed some light on DM models. Only the ultra-high energy cosmic ray observatories have the capabilities to detect such UHE decay products via the measurements of UHE photon induced extensive air showers. 
Using the upper bound on the flux of UHE cosmic rays beyond $10^{11.3}$ GeV implying $J(>10^{11.3}~{\rm{GeV}})< 3.6\times 10^{-5}$ km$^{-2}$sr$^{-1}$y$^{-1}$, at the $90\%$ C.L. reported by the Pierre Auger Observatory, we obtain global limits on the lifetime of the DM particles with masses $10^{15}\le M_{X} \le 10^{17}$ GeV. The constraints derived here are new and cover a region of the parameter space not yet explored. We compare our results with the projected constraints from future POEMMA and JEM-EUSO experiments, in order to quantify the improvement that will be obtained by these missions. 

Moreover, assuming that an epoch of early PBHs domination introduces a unique spectral break, $f_{\ast}$, in the gravitational wave spectrum, the frequency of which is related to the SHDM mass, we map potential probes and limits of the DM particles masses on the $f_{\ast}-M_{X}$ parameter space.

\end{abstract}

%\pacs{12.38.Gc, 12.38.Lg, 12.38.Aw}%PACS, the Physics and Astronomy %Classification Scheme.
%Use showkeys class option if keyword%display desired
\maketitle

%%%%%%%%%%%%%%%%%%%%%%%%%%%%%%%%
%%%%%%%%%%%%%%%%%%%%%%%%%%%%%%%%
\section{Introduction: Sources of Constraints on X-particles}
\label{sec:Intro}

The current cosmological understanding of structure formation, without taking the hazardous jump away from standard General Relativity gravity implies postulating the existence of Dark Matter (DM). 
Such existence, that entails specific gravitational consequences on structures that have been well established \citep{Betoule:2014frx,Ade:2013zuv}, although cosmologically dominant, continues to fail to produce constituent particles events in direct detection attempts, whether through nuclear recoil tanks or in accelerators \citep{Chatrchyan:2012me,ATLAS:2012ky,SuperCDMS:2014cds,Angloher:2011uu,Felizardo:2011uw,Klasen:2015uma,Akerib:2013tjd,Ahmed:2010wy,Bernabei:2010mq,CoGeNT:2010ols,XENON100:2012itz}, or in indirect surveys seeking detection through annihilation events of weakly interacting massive particle (WIMP) \citep{Conrad:2014tla}.

So far, many fundamental characteristics of DM particles remain quite unconstrained, including their mass or their self-annihilation time. Solutions to problems of the particle physics Standard Model (SM) have given birth to a wealth of DM models. Although most of them keep DM in a separate noninteracting sector, none of the models have won over the favors of the field because of their compelling theoretical appeal.

For many decades, the favored models characterized DM as a relic density of WIMPs \citep{Lee:1977ua,Vysotsky:1977pe,Goldberg:1983nd,Steigman:1984ac} (for 
a precise calculation of the WIMP relic abundance, see \citep{Gondolo:1990dk,Steigman:2012nb}; partial wave unitarity dictates an upper bound on the WIMP mass $\le 110$ TeV).
However, the extensive experimental program set up for WIMP detection in the direct and indirect detection experiments
\citep{Fermi-LAT:2012pls,
	MarrodanUndagoitia:2015veg,
	LUX:2016ggv,	
	Albert:2016emp,
	MAGIC:2016xys,
	XENON:2017vdw,
	PICO:2017tgi,	
	IceCube:2017rdn,
	PandaX-II:2017hlx,	
	HAWC:2017mfa,
	HAWC:2017udy,
	HESS:2018cbt,	
	Monteiro:2020wcb} as well as in the LHC has given negative or inconclusive results so far \citep{Buchmueller:2017qhf,Penning:2017tmb,Rappoccio:2018qxp}. 
Despite these facts, a complete exploration of the WIMP parameter space remains the highest priority of the DM community, and WIMP discovery is still a viable option in next-generation experiments. 

Because the assumption of relatively low-mass DM seems quite natural, it is rarely questioned. 

However, the null results from DM searches began closing the favored parameter space for the WIMP models and at the same time started to open a door to alternatives to the WIMP paradigm.

More recently, dark sector interactions have been given more serious consideration, which viability has been scrutinized in \citet{Comelli:2003cv}. Observational effects on galaxy clusters dynamics of such interactions have gathered credibility \cite{Bertolami:2007zm,Bertolami:2007tq,LeDelliou:2007am,Abdalla:2007rd,Bertolami:2008rz,Abdalla:2009mt,Bertolami:2012yp,LeDelliou:2014fto,LeDelliou:2018vua}. However, most of the current constraints tend to favour stable or long-lived, cold or warm, non-baryonic DM \cite{Bertone2005}. They consequently prompted the exploration of improved interacting models that respect those constraints, opening fairly large mass and interaction strength ranges for acceptable DM candidates \citep{Feng:2010gw}.

Alternative views on DM have emerged a while ago. The most radical of them suggests that DM interacts only gravitationally, explaining the negative experimental results. For example, 
cold DM could be a manifestation of the gravitational sector itself consisting of massive gravitons of bi-metric gravity \citep{Babichev:2016hir,Aoki:2016zgp} --- the only known self-consistent, ghost-free extension of General Relativity with massive spin-2 fields \citep{Hassan:2011zd}.

Among the well-motivated and less radical ideas for what DM could be, the WIMPzilla hypothesis postulates that DM is made of gravitationally produced (non-thermal relic) super weakly-interacting super massive so-called X-particles --- the Super-Heavy Dark Matter (SHDM) \citep{Chung:1998zb,Kuzmin:1998uv,Kuzmin:1998kk,Kolb:1998ki,Chung:1999ve,Kuzmin:1999zk,Chung:2001cb}. The hypothesis of DM consisting of heavy long-lived 
particles has attracted significant attention in the context of inflationary cosmology \citep{Kuzmin:1997jua,Berezinsky:1997hy,Kolb:2007vd,Kannike:2016jfs}. 
There are several scenarios of effective DM particles production at various stages of early Universe evolution. SHDM can be created gravitationally at the end of inflation \citep{Kuzmin:1998uv,Kuzmin:1998kk,Chung:1998zb,Kolb:2007vd}, during preheating \citep{Greene:1997ge,Chung:1998ua} and reheating \citep{Kuzmin:1997jua,Chung:1998rq,Gorbunov:2010bn}, and from the collisions of vacuum bubbles at phase transitions \citep{Chung:1998ua,Kolb:1998ki}, or from emission 
\citep{Samanta:2021mdm} of the primordial black holes (PBHs) \citep[see][for a review]{Kuzmin:1999zk}.

Of course, SHDM particles have been considered before to a certain extent. In particular, there is an extensive literature regarding observational constraints on unusually heavy DM candidates \citep[for example, see][and references therein]{DeRujula:1989fe, Ellis:1990nb, Sarkar:1995dd, Chianese:2021jke}. 

The earliest observational hint of SHDM existence was provided in the AGASA experiment by detection of super–GZK cosmic rays \citep[with GZK standing for Greisen, Zatzepin, and Kuzmin]{Takeda:2002at}. This GZK cut–off was later supported in results from next-generation cosmic ray experiments \citep{Abu_Zayyad_2013,PierreAuger:2008rol}. The IceCube PeV neutrinos detection \citep{IceCube:2013low,IceCube:2014stg} have lead to various DM decay propositions of interpretations \citep{Murase:2015gea,Bhattacharya:2014vwa,Esmaili:2013gha,Dev:2016qbd,Esmaili:2014rma,Rott:2014kfa}. However, recent analyses of the respective gamma-ray signal \citep{Kuznetsov:2016fjt,Cohen:2016uyg} have deprecated most of them, although the photon production suppression feature of a few of those DM models keep them conceivable \citep{Dev:2016qbd,Feldstein:2013kka,Hiroshima:2017hmy}. 
The search for signatures of Planckian-interacting massive particles in the data of the  Pierre Auger Observatory and derive constraints were reported in \citep{PierreAuger:2022ubv,PierreAuger:2022jyk}. 
The ultrahigh-energy cosmic rays (UHECRs) flux appears, from experimental data, to be dominated, above its ``ankle'', by an astrophysical, extra-galactic component \citep{PierreAuger:2017pzq}, most possibly sourced from starburst galaxies \citep{PierreAuger:2018qvk}. Supposing the systematic uncertainties of current high-energy hadronic interaction models remain sufficiently small, we can expect the acceleration of heavier nuclei, in addition to protons, as sources of this flux. 

Nevertheless, experimental data still allow for a minority component of different origin which, beyond the suppression, could still dominate the flux \citep{Alcantara:2019sco, Chianese:2021jke}.

The possibility that the by-products of the decay of unstable SHDM particles can contribute to the UHECR flux has been studied extensively in the past \citep[see for instance][]{Medina-Tanco:1999fld, Aloisio:2007bh, Kalashev:2007ph, Aloisio:2015lva, Kalashev:2017ijd, Marzola:2016hyt}. 
In these models, DM is composed of supermassive particles produced gravitationally during inflation \citep{Kuzmin:1998uv,Chung:1998zb,Kuzmin:1998kk,Kuzmin:1999zk}. These particles would be clustered in the halo of the galaxies, such as ours.

The discovery of gravitational waves (GWs) by LIGO and Virgo collaboration of black holes \citep{LIGOScientific:2016aoc,LIGOScientific:2016sjg,LIGOScientific:2017bnn,LIGOScientific:2017vox,LIGOScientific:2017ycc} and neutron stars
\citep{LIGOScientific:2017vwq,LIGOScientific:2020zkf} has opened up a new cosmic frontier for the SHDM search by examination of the stochastic GW background (SGWB) \citep{Maggiore:1999vm,Romano:2016dpx,Caprini:2018mtu,Christensen:2018iqi} in the multi-frequency range \citep{Samanta:2021mdm}.

Two main problems should be addressed in the discussion of SHDM models: how particles with very high mass ($M_{X}> 10^{13}$ GeV) can be quasi-stable, with a lifetime much longer than the age of the Universe $t_{0}$, and how their abundance can be dominant in the Universe today. The stability of SHDM can be achieved assuming the existence of a discrete gauge symmetry that protects the particle from decaying, in the same way as neutralino stability through $R$-parity in Super Symmetry. This discrete symmetry can be weakly broken, assuring a lifetime  $\tau_{X}>t_{0}$, through wormhole \citep{Berezinsky:1997hy} or instanton \citep{Kuzmin:1997jua} effects. 
An example of a particle with a lifetime exceeding the age of the Universe can be found in \citep{Ellis:1990iu}. 
Instanton decays induced by operators involving both the hidden sector and the SM sector may give rise to observable signals in the spectrum of UHECRs \citep{Berezinsky:1997hy,Kuzmin:1997jua}. Another possible way to stabilize such particles may be a modification of the standard cosmological expansion law in such a way that the density of these heavy relics would be significantly reduced \citep{Arbuzova:2020pka}.

Therefore, technically the SHDM models have two main parameters: mass $M_{X}$ and lifetime, $\tau_{X}$, in which a minority component of the UHECRs originates from the decay of these unstable particles. Stable X-particles are not so interesting from the experimental point of view since their annihilation cross-section is bounded by unitarity: $\sigma_{X}^{ann}\sim 1/M_{X}^{2}$, which makes its indirect detection impossible for today's experiments \citep{Gorbunov:2011zzc}. 
The spectrum from SHDM decay is expected to be dominated by gamma rays \citep{Kalashev:2016cre, Murase:2012xs,Esmaili:2015xpa} and neutrinos \citep{Murase:2015gea, Kachelriess:2018rty} because of more effective production of pions than nucleons in the QCD cascades. Since the photons would not be attenuated owing to their proximity, they become the prime signal because it is easier to detect photons than neutrinos. However, such gamma-ray production can be substantially different for different decay channels \cite {Chianese:2021jke}.

While it is very challenging to probe such superheavy DM via the traditional direct, indirect, or collider experiments, this paper aims to show the current bounds on such X particles, with a mass larger than the weak scale by several (perhaps many) orders of magnitude, from the perspective of the recent cosmic ray and GWs observations. In this paper, we examine particles' mass range from $10^{15}$ GeV to $10^{17}$ GeV together with the most recent UHECR and GWs data to derive the strongest lower limit on the lifetime of decaying superheavy WIMPs (WIMPzillas) and their mass. 
In Sec.~\ref{sec:GammaRayFlux_from_SHDM} we estimated the photon flux from SHDM decays. For this, we evaluated two separate contributions: the astrophysical factor and the particle physics factor. We used current high-energy gamma-ray measurements \cite{Veberic:2017hwu} to examine bounds that one may put on the parameter space of decaying SHDM. 
In addition, from the relation of the GW spectrum break, $f_{\ast}$, with the SHDM mass, following Ref.~\citep{Samanta:2021mdm}, we mapped potential probes and limits of the SHDM particles masses on the $f_{\ast}-M_{X}$ parameter space in Sec.~\ref{sec:BoundsSHDM_GW}. 
Result discussion is given in Sec.~\ref{sec:Results} and we 
conclude in Sec.~\ref{sec:Conclusion}.

%%%%%%%%%%%%%%%%%%%%%%%%%%%%%%%%
%%%%%%%%%%%%%%%%%%%%%%%%%%%%%%%%
\section{Bounds on SHDM from the gamma-ray flux}
\label{sec:GammaRayFlux_from_SHDM}

Given the mass of the SHDM particles, the decay time corresponding to the scenario for the largest SHDM cosmic ray flux, compatible with the upper limits to the photon fraction
obtained by the Pierre Auger Observatory (Auger), can be estimated from the predicted integral gamma-ray flux \cite{Kalashev:2016cre}, which may be observed on Earth by an 
observatory with uniform exposure, as
\begin{equation}
	J(E)  = \frac{N(E>E_{\rm{min}})}{4\pi M_{X}\tau_{X}} \Bigg \{	\frac{\int_{V}^{} \frac{\rho_{DM}(r)\omega(\delta,a_{0},\theta_{\rm{max}})}{r^{2}} ~dV}{2\pi \int_{-\frac{\pi}{2}}^{\frac{\pi}{2}} \omega(\delta, a_{0}, \theta_{\rm{max}}) {\rm{cos}}(\delta) ~d\delta}	 \Bigg \}.
	\label{eq:CR_flux_GalaxyUnit_volume}
\end{equation}
This expression can be easily adopted in case we start to look at cosmic rays using the Moon's regolith \citep{Romero-Wolf:2021wbw}. 
Here $N(E > E_{\rm{min}})$ is an integral number of photons with energies higher than $E_{\rm{min}}$ produced in the decay of X-particle; $\theta$ is the angle between the line of sight and the axis defined by Earth and the Galactic center \citep{Aloisio:2006yi}; $\tau_{X}$ is the SHDM lifetime, $10^{11}-10^{22}$ years;
$M_{X}$ is the SHDM particle mass, $10^{15}-10^{17}$ GeV; 
$\rho_{DM}(r)\equiv\rho_{X}(r)$ is the density of 
DM in the Galaxy as function of distance, $r$, from the Galactic Center, in $\rm{\frac{GeV}{cm^3}}$. 
Integration in the numerator ranges
over all the volume of the halo ($R_{H} = 260$ kpc) and in the denominator over all the sky (the averaging over right ascension is included in the definition of the directional exposure, $\omega(\hat{\bf{n}})$).

The directional exposure, $\omega(\hat{\bf{n}})$, provides the effective time-integrated collecting area for a flux from each direction of the sky $\hat{\bf{n}}(\alpha, \delta)$, characterized by the right ascension $\alpha$ and the declination $\delta$. For an experiment at latitude $\lambda$, which is fully efficient for particles arriving with zenith angle $< \theta_{\rm{max}}$ and that experiences stable operation, $\omega(\hat{\bf{n}})$ actually becomes independent of $\alpha$ when integrating the local-angle-detection efficiency over full periods of sidereal revolution of the Earth. Full efficiency means that the acceptance depends on $\theta$ only through the reduction in the perpendicular area given by ${\rm{cos}}(\theta)$. 

The $\omega$ dependence on declination, $\delta$, geographical latitude of the given experiment $\lambda$ and the maximal zenith angle $\theta_{\rm{max}}$ accessible for fully efficient observation in the experiment, 
relies on geometrical acceptance terms and is given by \cite{Sommers:2000us, TelescopeArray:2014ahm}
\begin{equation}
\begin{split}
\omega(S,\Delta t,\delta,\lambda,\theta)  = &\hphantom{+.}\frac{S\Delta t}{2\pi}  {\rm{cos}}(\lambda){\rm{cos}}(\delta){\rm{sin}}(a_{m}(\theta,\delta,\lambda)) \\
& + \frac{S\Delta t}{2\pi} a_{m}(\theta,\delta,\lambda){\rm{sin}}(\lambda){\rm{sin}}(\delta)   ,
\end{split}		
\end{equation}
where $S$ is the effective surface of the given experiment (detector array); $\Delta t$ is the total exposure time of the given experiment (or time of data collection). 
The location of the Pierre Auger Observatory, $35.1^{\circ}-35.5^{\circ}S, 69.6^{\circ}W$ at 875 g/cm$^{2}$ atmospheric depth \citep{ABRAHAM200450}, corresponds to the Pampa Amarilla plain, in the Mendoza Province of Argentina, close to the Malarg\"{u}e town. 
Initiating data collection in January 2004 and completing its baseline design construction by 2008, by 2021, the Auger collected exposure had exceeded $7.68 \times 10^{4}$km$^{2}$ sr yr, exceeding the sum over all of the other cosmic ray experiments available \citep{Miramonti_2021}. Therefore, for the Auger with $S=1.037\times 10^{4}$ km$^{2}$, 
$\Delta t=7.41$ years from Ref.~\citep{Miramonti_2021},
$\lambda = -35.2^{\circ}$ and
$-15^{\circ} \le \delta \le 25^{\circ}$. 
	
The parameter $a_{m}$ of the observatory is given by
\[
a_{m}(\theta,\delta,\lambda) =
\begin{cases}
	0   &\text{for~~} \xi\textgreater 1\\
	\pi &\text{for~~} \xi\textless -1\\
	\rm{arccos}(\xi) &\text{otherwise},
\end{cases}
\]
with
\begin{equation}
\xi(\theta,\delta,\lambda)= \frac{{\rm{cos}}(\theta_{\rm{max}})- {\rm{sin}}(\lambda){\rm{sin}}(\delta)}{{\rm{cos}}(\lambda){\rm{cos}}(\delta)}.
\end{equation}

The DM galactic distribution, or DM density profile $\rho_{DM}$, is a function of the
Galactic longitude, $l$, and latitude $b$, which in turn is related to the line-of-sight, $s$, coordinate:
\begin{equation}
	r(s,b,l)=\sqrt{s^{2}+r_{\odot}^{2}-2sr_{\odot}{\rm{cos}}(b){\rm{cos}}(l)},
\end{equation}
where $r_{\odot}=8.5$ ${\rm{kpc}}$ denotes the distance between the Earth and the Galactic Center. The local DM density is an important ancillary parameter when constraining DM signatures. However due to the lack of a robust estimate of $\rho_{X}$, the distribution of DM is assumed to follow a density profile inspired by numerical simulations, typically an analytic fit such as the well-known Navarro-Frenk-White (NFW) \citep{Navarro:1995iw} or
Einasto \cite{Einasto:1965czb, Graham:2005xx} profiles, with two or more free parameters whose best-fit values are then determined from dynamical constraints. 

In this work we adopt both reference profiles: the NFW profile 
\begin{equation}
	\rho^{\rm{NFW}}_{X}(r) = \rho_{s} \left[ \frac{r}{r_{s}}\left(1 + \frac{r}{r_{s}}\right)^{2} \right]^{-1} 
\end{equation}
with $r_{s} = 28.44$ kpc \citep{Navarro:1995iw}; the Einasto profile 
\begin{equation}
	\rho^{\rm{Einasto}}_{X}(r)=\rho_{s} e^{-\frac{2}{\alpha} \left[ \left( \frac{r}{r_{s}}\right)^{\alpha}-1 \right] },
\end{equation}
in contrast with the somewhat steeper NFW profile, with $r_{s} = 30.28$ kpc, 
$\alpha= 0.17$ \citep{Einasto:1965czb, Navarro:2008kc}
and $\rho_{s} = 0.105$ GeV cm$^{-3}$ \cite{Alcantara:2019sco}. 
Most recent systematic efforts to estimate a proper DM profile were summarized in Ref.\cite{deSalas:2020hbh}.

With all this in hand, and defining the element of Galaxy volume in Eq.~\eqref{eq:CR_flux_GalaxyUnit_volume} as
\begin{equation}
	dV=r^{2}{\rm{sin}}(\varphi)drd\varphi d\theta=r^{2}{\rm{sin}}\left(\frac{\pi}{2}-\delta\right)~drd\delta d\theta 
\end{equation}
one can rewrite the flux expression \eqref{eq:CR_flux_GalaxyUnit_volume}, as following
\begin{widetext}
	\begin{equation}
		J(E)  =  \frac{1}{4\pi}\frac{1}{M_{X}\tau_{X}}\frac{dN}{dE} \Bigg \{ \frac{\int_{0}^{50{\rm{kpc}}} \frac{\rho_{X}(r)}{r^{2}}r^{2}dr \int_{-\frac{\pi}{2}}^{\frac{\pi}{2}} {\rm{sin}}(\frac{\pi}{2}-\delta)~d\delta \int_{0}^{2\pi}   \frac{\omega(\delta,a_{0},\theta_{\rm{max}})}{1} ~d\theta}{2\pi \int_{-\frac{\pi}{2}}^{\frac{\pi}{2}} \omega(\delta, a_{0}, \theta_{\rm{max}}) {\rm{cos}}(\delta)~d\delta} \Bigg \}.
		\label{eq:SHDM_CRflux}
	\end{equation}
\end{widetext}
Using the radius of the Galactic halo $R_{H}=260~{\rm{kpc}}$, the integral in Eq.~\eqref{eq:SHDM_CRflux} can be splitted in two integrals as was shown in Refs.~\citep{Supanitsky:2019ayx, Alcantara:2019sco}.
\begin{widetext}
	\begin{equation}
		J(E,\theta) = \frac{1}{4\pi}\frac{1}{M_{X}\tau_{X}}\frac{dN}{dE} \Bigg \{2 \int_{r_{\odot}{\rm{sin}}\theta}^{r_{\odot}} r \frac{\rho_{X}(r)}{\sqrt{r^{2}-r^{2}_{\odot}{\rm{sin}}^{2}\theta}} dr  + \int_{r_{\odot}}^{R_{H}} r \frac{\rho_{X}(r)}{\sqrt{r^{2}-r^{2}_{\odot}{\rm{sin}}^{2}\theta}} dr \Bigg \},
		\label{eq:DMflux}
	\end{equation}
\end{widetext}
Therefore, the expected energy distribution on Earth follows the initial decay spectrum, whereas the angular distribution incorporates the (uncertain) distribution of DM in the Galactic halo via the line-of-sight integral \citep{Dubovsky:1998pu, Evans:2001rv, Aloisio:2007bh, Kalashev:2017ijd}.

\begin{figure*}[t]
	\centering
	\includegraphics[width=0.9\textwidth]{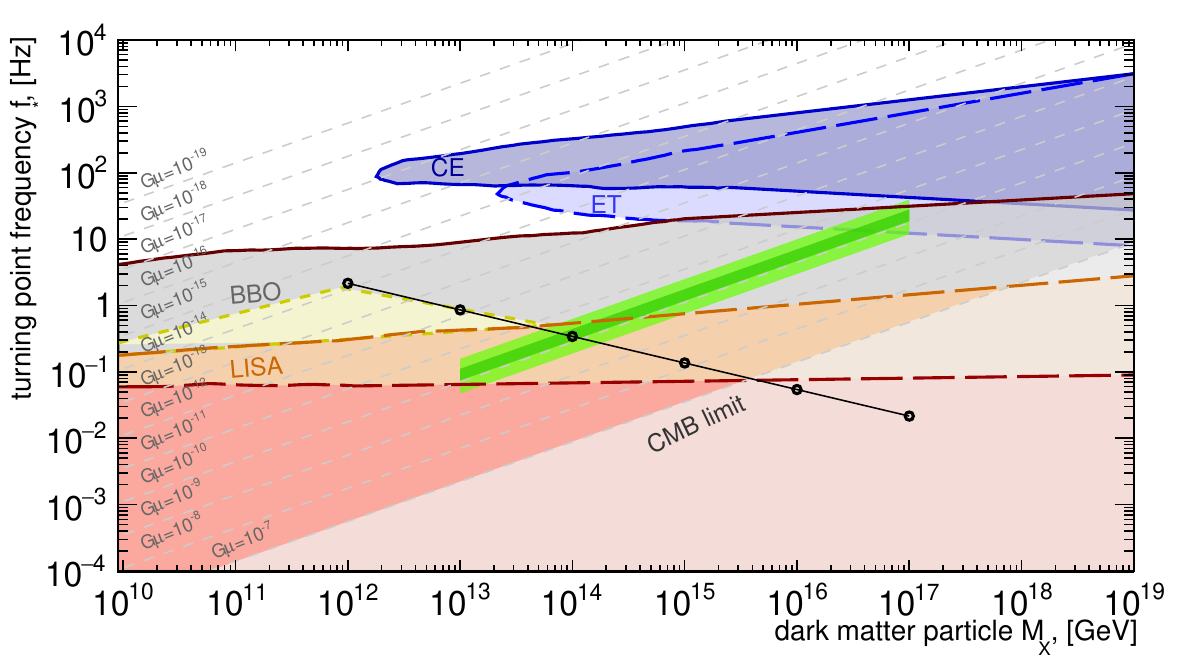}
	\caption{
		Map of the possible probes of $M_{X}$ and hence of 
		the PBH evaporation temperature projecting the turning-point frequencies $f_{\ast}$, Eq.~\eqref{eq:tp_frequency}, for the various SHDM particle masses.
		Here, each diagonal line corresponds to the $G\mu$ value in the range $10^{-19}$ to $10^{-7}$ running from top to bottom. Below  $G\mu=10^{-7}$ lies 
		the CMB limit for the cosmic string tension \cite{Torki:2021fvi}. 
		The relevant sensitivity curves of currfnt or future GW detectors, such as the Einstein Telescope (ET), Cosmic Explorer (CE), LISA\citep{LISA:2017pwj}, BBO \citep{Corbin:2005ny} are reproduced from \citep{Samanta:2021mdm} using 
		Eq.~\eqref{eq:tp_frequency}. 
		The light yellow region is the allowed parameter space assuming the PHB emission to SHDM scenario from \citep{Samanta:2021mdm}.
		The dark and light green bands denote the predicted turning-point frequencies computed assuming the 1$\sigma$ and 2$\sigma$ prediction on $G\mu$ parameter taken from \citep{Blasi:2020mfx} obtained from the fits to the NANOGrav data \citep{NANOGrav:2020bcs} for the examined SHDM particle masses.
	}
	\label{fig:turnPointFreq_Mdm_constraints}
\end{figure*}

The energy spectrum, $dN/dE$, of the expected gamma rays depends on the exact SHDM decay mechanism and is model dependent. Trying to avoid this issue, in our computation we used the following flat spectrum similar to \cite{Alcantara:2019sco}, 
\begin{equation}
\frac{dN\left(E \right)}{dE}  \sim {\rm{cos}}(\delta) \omega(S, \Delta t,\delta, \lambda,\theta) \left[M_{X}^{0.9} {\rm{log}}\left(\frac{2E}{M_{X}}\right)^{-1.9} \right],
\label{eq:DMflux_PAO}
\end{equation}
validated for $X \rightarrow q\bar{q}$ decay, it is independent of the particle
type, assuming the photon/nucleon ratio is $2 \lesssim \gamma/N \lesssim 3$
and the neutrino/nucleon ratio is $3 \lesssim \nu/N \lesssim 4$. 
In comparison with the other spectra 
available in the literature \cite{Bergstrom:1997fj,Tasitsiomi:2002vh, Bringmann:2007nk, Calore:2016ogv}, the spectrum from \cite{Alcantara:2019sco} is much flatter. This allows one to have fixed photon/nucleon and neutrino/nucleon ratios along the whole energy range.

The consideration of the bosonic decay channels such as WW/ZZ/hh, or leptonic ee/$\mu\mu$/$\tau\tau$ would require a different stable particle spectra in Eq.~\eqref{eq:DMflux}.

The Auger is composed of two types of instruments: \begin{inparaenum}[\itshape a\upshape)]
                          \item fluorescence telescopes, that measure light from atmospheric nitrogen excitation by air shower particles,
                          \item ground particle detectors, that sample air shower fronts arriving at the Earth’s surface.
             \end{inparaenum} Its maximum zenith angle falls at $\theta_{\rm{max}}=90^{\circ}$, with the downward-going (DG) channel constrained to $60^{\circ} \le \theta \le 90^{\circ}$ while the Earth-skimming (ES) channel extends to $60^{\circ} \le \theta \le 95^{\circ}$ \citep[see][]{Pedreira:2021gcl}.          
Therefore, in order to make a direct comparison with current limits on the diffuse UHE gamma-ray flux, from Eq.~\eqref{eq:DMflux}, following \citep{Kalashev:2016cre}, we compute the angle-averaged integral $\gamma$-flux over the whole sky $(0 < \theta < \pi)$, averaging over the directional exposure at the declination of the Auger Observatory, where 
the declination limits are $-15^{\circ} \le \delta \le 25^{\circ}$ \citep{TelescopeArray:2014ahm}.

The SHDM flux contributions from the extragalactic and galactic haloes need to be resolved as they 
can be important due to the fact that the gamma rays and protons originating in that extragalactic halo come from a narrow region of the sky. 
Therefore, in that region the contribution of the extragalactic halo can be more important than the one corresponding to our galaxy, specially in regions far from the galactic center where the galactic contribution decreases considerably \citep{Supanitsky:2019ayx}. 

The fluxes obtained for the various DM mass and DM density profiles are listed in Table.~\ref{tab:DM_fluxes}.

\begin{table}[t]
	\begin{tabular}{l|ll|ll}
		\multirow{1}{*}{$M_{X}$,} 
		&\multirow{1}{*}{$\tau_{X}$,} 
		&\multirow{2}{*}{$J(> E_{0})$} 			
		&\multirow{1}{*}{$\tau_{X}$,} 
		&\multirow{2}{*}{$J(> E_{0})$} \\		
		\multirow{1}{*}{[GeV]} 		
		&\multirow{1}{*}{[years]} 
		&\multirow{1}{*}{} 			
		&\multirow{1}{*}{[years]}
		&\multirow{1}{*}{} \\	
		\hline		
		\multicolumn{5}{c}{Einasto, $r_{s}=30.28$ kpc} \\
		\hline
		$10^{15}$ &$10^{21}$ &$6.803 \times 10^{-5}$ &$10^{22}$ &$6.803 \times 10^{-6}$ \\
		$10^{16}$ &$10^{21}$ &$7.406 \times 10^{-5}$ &$10^{22}$ &$7.406 \times 10^{-6}$ \\
		$10^{17}$ &$10^{21}$ &$8.571 \times 10^{-5}$ &$10^{22}$ &$8.571 \times 10^{-6}$ \\ 		
		\hline		
		$10^{15}$ \cite{Bergstrom:1997fj} &$10^{21}$ &$1.028 \times 10^{-3}$ &$10^{22}$ &$1.028 \times 10^{-4}$ \\ 
		$10^{16}$ \cite{Bergstrom:1997fj} &$10^{21}$ &$3.250 \times 10^{-3}$ &$10^{22}$ &$3.250 \times 10^{-4}$ \\	
		$10^{17}$ \cite{Bergstrom:1997fj} &$10^{21}$ &$1.028 \times 10^{-2}$ &$10^{22}$ &$1.028 \times 10^{-3}$ \\
		\hline				
		$10^{15}$ \cite{Tasitsiomi:2002vh} &$10^{21}$ &$5.867 \times 10^{-4}$ &$10^{22}$ &$5.867 \times 10^{-5}$ \\
		$10^{16}$ \cite{Tasitsiomi:2002vh} &$10^{21}$ &$1.855 \times 10^{-3}$ &$10^{22}$ &$1.855 \times 10^{-4}$ \\	
		$10^{17}$ \cite{Tasitsiomi:2002vh} &$10^{21}$ &$5.867 \times 10^{-3}$ &$10^{22}$ &$5.867 \times 10^{-4}$ \\					
		\hline		
		\multicolumn{5}{c}{NFW, $r_{s}=28.44$ kpc} \\
		\hline
		$10^{15}$ &$10^{21}$ &$7.736 \times 10^{-5}$ &$10^{22}$ &$7.736 \times 10^{-6}$ \\
		$10^{16}$ &$10^{21}$ &$8.112 \times 10^{-5}$ &$10^{22}$ &$8.112 \times 10^{-5}$ \\
		$10^{17}$ &$10^{21}$ &$8.186 \times 10^{-5}$ &$10^{22}$ &$8.186 \times 10^{-5}$\\				
		\hline		
		$10^{15}$ \cite{Bergstrom:1997fj} &$10^{21}$ &$2.047 \times 10^{-4}$ &$10^{22}$ &$2.047 \times 10^{-5}$ \\
		$10^{16}$ \cite{Bergstrom:1997fj} &$10^{21}$ &$6.473 \times 10^{-4}$ &$10^{22}$ &$6.473 \times 10^{-5}$ \\	
		$10^{17}$ \cite{Bergstrom:1997fj} &$10^{21}$ &$2.047 \times 10^{-3}$ &$10^{22}$ &$2.047 \times 10^{-4}$\\
		\hline				
		$10^{15}$ \cite{Tasitsiomi:2002vh} &$10^{21}$ &$1.168 \times 10^{-4}$ &$10^{22}$ &$1.168 \times 10^{-5}$ \\
		$10^{16}$ \cite{Tasitsiomi:2002vh} &$10^{21}$ &$3.694 \times 10^{-4}$ &$10^{22}$ &$3.694 \times 10^{-5}$ \\	
		$10^{17}$ \cite{Tasitsiomi:2002vh} &$10^{21}$ &$1.168 \times 10^{-3}$ &$10^{22}$ &$1.168 \times 10^{-4}$\\
		\hline		
		\multicolumn{5}{c}{NFW, $r_{s}=30.28$ kpc} \\
		\hline		
		$10^{15}$ &$10^{21}$ &$5.451 \times 10^{-5}$ &$10^{22}$ &$5.451 \times 10^{-6}$ \\
		$10^{16}$ &$10^{21}$ &$5.675 \times 10^{-5}$ &$10^{22}$ &$5.675 \times 10^{-6}$ \\		
		$10^{17}$ &$10^{21}$ &$5.988 \times 10^{-5}$ &$10^{22}$ &$5.988 \times 10^{-6}$\\						    		
		\hline				
	\end{tabular}
	\caption{
		Integral fluxes for the DM particle candidates with mass, $M_{X}$. DM fluxes computed with the help of Eq.~\eqref{eq:DMflux} assuming the Einasto and NFW DM density profiles. In addition, we show results obtained with the energy spectra from \cite{Bergstrom:1997fj,Tasitsiomi:2002vh}.
 }
\label{tab:DM_fluxes}
\end{table}

%%%%%%%%%%%%%%%%%%%%%%%%%%%%%%%%
%%%%%%%%%%%%%%%%%%%%%%%%%%%%%%%%
\section{Bounds on SHDM from the Gravitational Waves Observations}
\label{sec:BoundsSHDM_GW}

Today the scientific community started to synthesize the various astronomical messengers, namely photons, neutrinos, cosmic rays, and gravitational waves. While photons, neutrinos, and cosmic rays still have a vital role to play in multi-messenger astronomy, 
GWs observation with pan-spectral electromagnetic radiation has enriched our understanding of violent astronomical events \cite{LIGOScientific:2017vwq}.

The details of the SHDM production mechanism can leave their footprint on the primordial GW amplitude as well as the spectral features. 
However, such a scenario is only present if DM mass would be generated by spontaneous symmetry breaking of an Abelian symmetry, which is not the general case.

An early Universe cosmological phase transition induced by this spontaneous symmetry breaking \citep{Mazumdar:2018dfl} may result into emergence of a cosmic string network \citep{Kibble_1976,Jeannerot:2003qv} -- line-like topological defects. 
Cosmic strings restore that broken Abelian symmetry at the core of these topological defects with vortex types of behavior \citep{NIELSEN197345}. Their network loses energy through the shrinking of its closed loops following their emission of GWs \cite{PhysRevD.31.3052,Auclair:2019wcv}. The resulting primordial GW signal contains key signatures of ultraviolet physics that would otherwise remain far beyond the reach of regular ground detection. This is why such signal is a main focus of current and future investigations of the stochastic GW background (SGWB) \citep{Maggiore:1999vm,Romano:2016dpx,Caprini:2018mtu,Christensen:2018iqi}. 
In addition, observation of such a signal \citep{Lentati:2015qwp, NANOGRAV:2018hou,NANOGrav:2020bcs} can be a complementary probe to the range of SHDM mass if one assumes that such DM particles are produced by emission of the PBHs.

The shape of the GW spectrum from a cosmic string network is expected to follow a convex cored power law which slope varies with amplitude and frequency, as well as is parameterized by the product $G\mu$ between the string tension $\mu$  and Newton's gravitation constant $G$.

% Figure
\begin{figure*}[t]
	\centering	
	\includegraphics[width=0.9\textwidth]{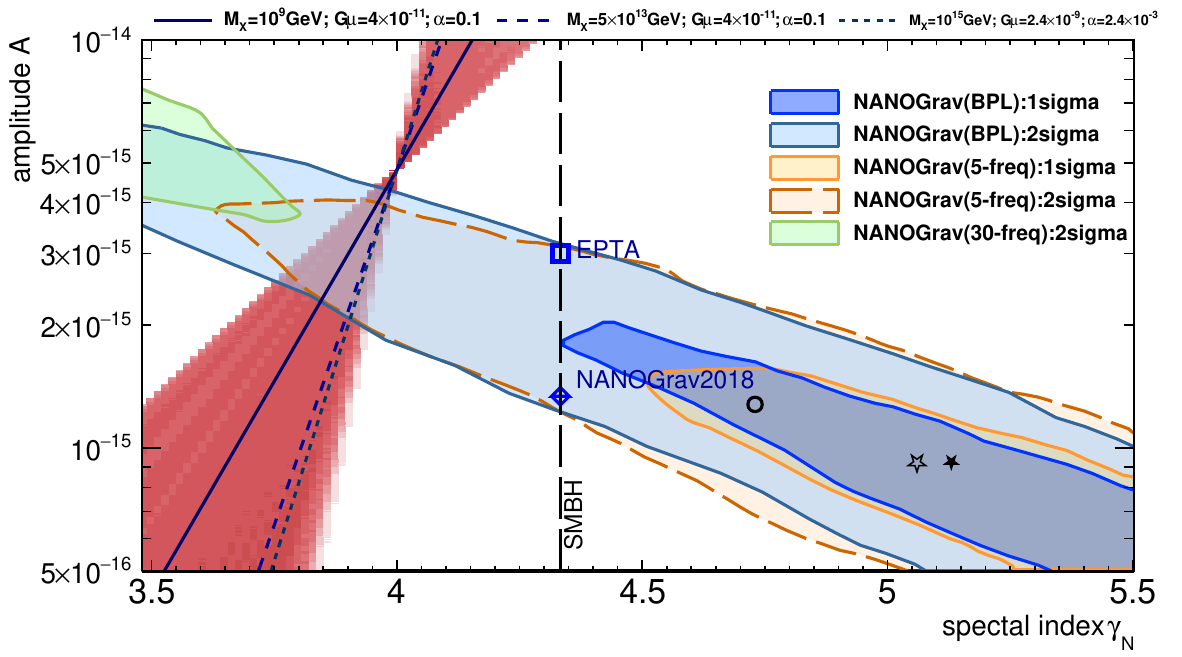}			
	\caption{
		The red graded area represents 
		solutions of Eq.~\eqref{eq:GW_AmplEquation} with %the 
		respect to the characteristic GW strain amplitude, $A$, at $f = f_{\ast}(M_{X})$, projected  onto the $\gamma_{N}-A$ plane. 
		Three benchmark points for $A$ and $\gamma_{N}$ parameters, which were listed in Ref.\citep{Blasi:2020mfx} are shown as markers ($\circ$, $\star$, $\smwhitestar$).		
		The dark and light contours denote the 1$\sigma$ and 2$\sigma$ posteriors in the NANOGrav analysis that allow to describe the observed stochastic process. Here, we use the contours taken from Ref.~\citep{NANOGrav:2020bcs}.
		The black dashed vertical line indicates the theoretical prediction for a population of
		supermassive black hole binaries (SMBHBs), $\gamma_{N} = 13/3$. 		
		Points on the vertical line denote a 95\% upper limit on the dimensionless strain amplitude %$A$ of 
		$A=3\times 10^{-15}$ and $A=1.34\times 10^{-15}$ at a reference frequency of 1 ${\rm{yr}}^{-1}$ and a spectral index of 13/3 obtained from the European Pulsar Timing Array (EPTA) data and from the NANOGrav report respectively \citep{Lentati:2015qwp, NANOGRAV:2018hou}.	
	}
	\label{fig:NANOGrav_comp}
\end{figure*}

The standard form of the spectral GWs energy density can be expressed today as the power-law:
\begin{equation}
	\label{eq:GWs_abundance}
	\Omega_{GW}(f)=\frac{2\pi^{2}}{3H_{0}^{2}}A^{2}f_{yr}^{2} \left( \frac{f}{f_{yr}} \right)^{5-\gamma_{N}},
\end{equation}
where $f$ is the oscillation frequency of the emitted GWs, $f_{yr}=1{\rm{yr}}^{-1}$, $A$ is the characteristic GW strain amplitude, $\gamma_{N}$ is the spectral index of the pulsar timing-residual cross-power spectrum and $H_{0}=67.7$ $\rm{{km}/({Mpc~sec})}$ is the present Hubble constant.

Following the recipe defined in \citep{Samanta:2021mdm} one can approximately determine the frequency at which the GW spectrum, $\Omega_{GW}(f)h^{2}$, changes slope from a plateau described by $f^{0}$ to $f^{-1/3}$ with the help of the following expression 
\begin{equation}
	\label{eq:tp_frequency}
	f_{\ast} \simeq 2.1\times 10^{-8} \sqrt{\frac{50}{z_{eq}\alpha \Gamma G\mu }} \left(\frac{M_{X}}{T_{0}} \right)^{\frac{3}{5}} T_{0}^{-\frac{2}{5}} t_{0}^{-1},
\end{equation}
where $\Gamma \simeq 50$ is the constant rate of GW radiation, $\alpha =0.1$ characterises 
the cosmic string loop size at the time of formation, $z_{eq} =3387$ is the red-shift at the usual matter-radiation equality, occurring 
at the time
$t_{eq}$, 
$T_{0} =2.725$ K is the photon temperature today, and $t_{0} = 13.81$ Gyr is the age of the Universe.

With Eq.~\eqref{eq:tp_frequency} we computed the turning-point frequencies, $f_{\ast}$, 
for the range of masses $M_{X}=10^{12}- 10^{17}$ GeV assuming 
the dimensionless combination
\citep{Samanta:2021mdm,Bian:2021vmi} between the corresponding cosmic string tension and SHDM particle mass,
\begin{equation}
G\mu \sim GM^{2}_{X}.
\end{equation}
The obtained frequencies lie within the range from 
2.16 to 0.021 Hz, for masses ranging from $10^{12}$ to $10^{17}$ GeV respectively. With such an assumption, furthermore admitting that the $G\mu$ value for 
$M_{X}=10^{16}$ GeV is below the Cosmic Microwave Background (CMB) limit on $G\mu$ \citep{Blasi:2020mfx}(see black data-points on Fig.~\ref{fig:turnPointFreq_Mdm_constraints}) while 
taking the resulted $1\sigma$ value of $G\mu$ \citep{Blasi:2020mfx} from the fits to the NANOGrav signal \citep{NANOGrav:2020bcs}, which is $[4,9]\times 10^{-11}$, we determined 
$f_{\ast}$ in the range of 1.76(1.17) to 27.94(18.63) Hz. This result is shown with the green band on Fig.~\ref{fig:turnPointFreq_Mdm_constraints} with the different probes of the $M_{X}$ and hence the PBH evaporation temperature. 
Assuming that PBHs evaporation causes a break in the GW spectrum at the turning point frequency \citep{Samanta:2021mdm}, $f_{\ast}(M_{X})$, from Eqs.~\eqref{eq:GWs_abundance} and \eqref{eq:tp_frequency} one may solve the following equality
\begin{equation}
	\label{eq:GW_AmplEquation}
\scalemath{0.8}{	
	2.1\times 10^{-8} \sqrt{\frac{50}{z_{eq}\alpha \Gamma G\mu }} \left(\frac{M_{X}}{T_{0}} \right)^{\frac{3}{5}} T_{0}^{-\frac{2}{5}} t_{0}^{-1} = \frac{2\pi^{2}}{3H_{0}^{2}}A^{2}f_{yr}^{2} \left( \frac{f_{\ast}(M_{X})}{f_{yr}} \right)^{5-\gamma_{N}}},
\end{equation}
with respect to the characteristic GW strain amplitude, $A$ by scanning over $\gamma_{N}$ in the range of 3.5 to 5.5 with the step size of 0.005, $G\mu$ in the range of $10^{-12}$ to $10^{-7}$ with the step size of $10^{-13}$, and $\alpha$ in the range of $10^{-6}$ to $0.2$ with the step size of $0.025$. 
The obtained solutions are compared with the NANOGrav observation in Fig.~\ref{fig:NANOGrav_comp}, where 
the red cone represents the domain of solutions for the mass range of 
$M_{X}=10^{9} - 10^{19}$ GeV. The results for the benchmark values of $G\mu$ and $\alpha$ parameters, which are listed in \citep{Blasi:2020mfx}, are shown with the solid and dashed blue lines to guide the eye.

%%%%%%%%%%%%%%%%%%%%%%%%%%%%%%%%
%%%%%%%%%%%%%%%%%%%%%%%%%%%%%%%%
\section{Results and Discussion}
\label{sec:Results}

% Figure
\begin{figure*}%[h]
	\centering
	\includegraphics[scale=0.85]{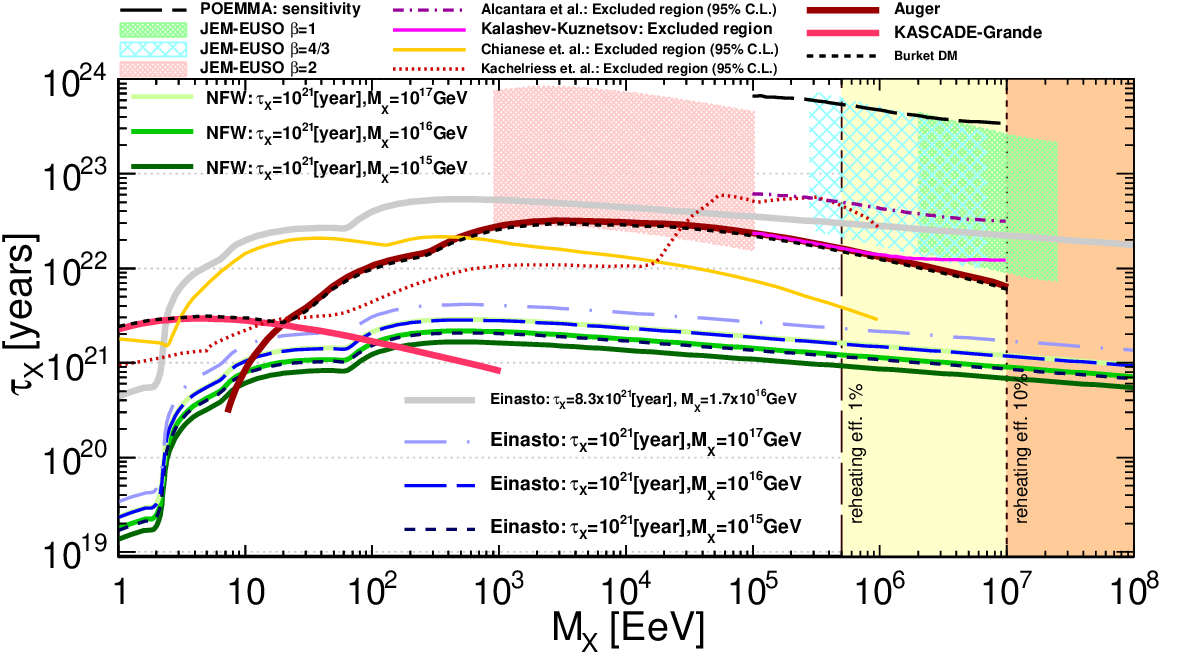}
	\includegraphics[scale=0.85]{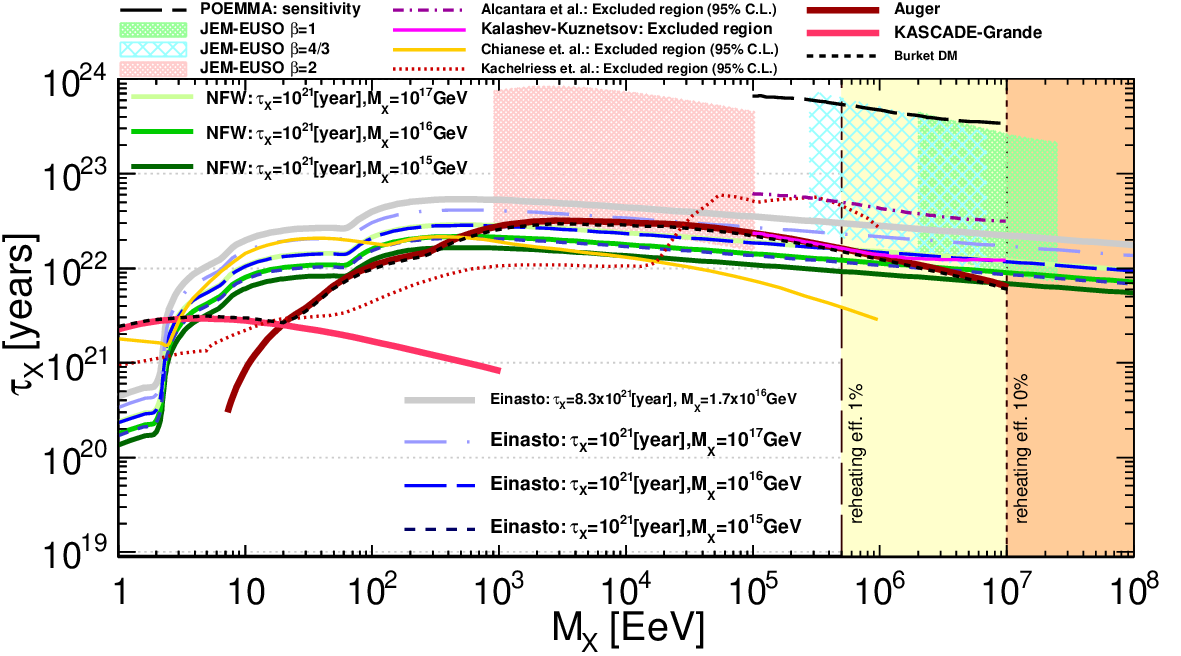}	
	\caption{
		Comparison of the upper constraints obtained in the literature with the 95\% C.L. exclusion plot for mass $M_{X}$ and lifetime $\tau_{X}$ of DM particles. 
		The constraints that were obtained with the data of Pierre Auger full-sky analysis \citep{Bister_2021} assuming NFW DM profile (solid green lines), or with assumption that the DM profile is given 
		by the Einasto model (dashed blue lines), 
		constraints obtained using the DM flux estimation from Mikhail Kuznetsov given in the Ref.\citep{Alcantara:2019sco}, are shown with the gray bold solid line.
		The lower limit on the lifetime of SHDM particles together with the stereoscopic $\tau_{x}$ sensitivity (defined by the observation of one photon event above $10^{11.3}$ GeV in 5 yr of data collection) 
		of POEMMA (dashed dark gray line) was taken from \citep{Alcantara:2019sco}. 	
		The Alcantara  95\% C.L. excluded region (dash-dotted dark magenta line) 
		and the Kalashed-Kuznetsov excluded region (solid magenta line), all those curves are taken from \citep{Alcantara:2019sco}. 		
		The regions accessible to the JEM-EUSO experiment, each region corresponding 
		to a different choice of the power law index in the inflation potential, 
		$\beta=2;4/3;1$, are plotted from left to right \citep{Aloisio:2015lva}.
		The gamma-ray limits placed by 
		\citet{Chianese:2021jke} (orange solid line), the limits placed by 
		\citet{Kachelriess:2018rty} (red dotted line) are also shown. 
		For illustration purposes, the 95\% CL upper limit on mass obtained from the possible value of the Hubble rate at the end of inflation for a reheating efficiency of 1\% (10\%) is shown as the vertical dashed (dotted) line \citep{Garny:2015sjg}. 
		The constraints derived from diffuse $\gamma$-ray and neutrino limits
		from the Auger (solid bold dark red line), from the KASKADE-Grande (solid bold red line). 						
		We also show for comparison the constraints obtained assuming Burkert DM profile (black short dashed line) using the data of Pierre Auger partial-sky analysis \citep{Bister_2021}.
	}
	\label{fig:DM_constraints}
\end{figure*}

The integral fluxes of photons at the location of the Auger Observatory for the different masses of the DM particle candidates are shown in Table~\ref{tab:DM_fluxes}. 

In Fig.~\ref{fig:DM_constraints} we compare the result of this work with the current lifetime limits placed by Auger, the KASCADE-Grande observations \citep{Alcantara:2019sco}. We add the existing gamma-ray limits  placed by 
\citet{Chianese:2021jke} (orange solid line) and 
\citet{Kachelriess:2018rty} (red dotted line). 
We also project the sensitivity regions for the future POEMMA,  JEM-EUSO space missions. The regions accessible by the POEMMA experiment are 
shown with the dashed black line \citep{Alcantara:2019sco}. The regions accessible to the JEM-EUSO experiment, each region corresponding 
to 
a different choice of the power law index in the inflation potential, $\beta=2;4/3;1$, are plotted from left to right \citep{Aloisio:2015lva}.

We see that the integral DM flux for the examined range of the DM particle candidate masses is of order $10^{-5}$, see Table \ref{tab:DM_fluxes}. With 
such a DM flux, the examined SHDM particle candidates with 
masses of $10^{15}-10^{17}$ GeV and lifetime of $10^{21}$ years may be already excluded by the current gamma rays observations, see Fig.~\ref{fig:DM_constraints} (top). In the case that  
some mechanisms could increase the DM flux by order of magnitude with the same lifetime, 
small 
regions on the $\tau_{X}-M_{X}$ parameter space could then 
be constrained, but only by 
future detectors, such as JEM-EUSO, POEMMA, see Fig.~\ref{fig:DM_constraints} (bottom).  
The constraints obtained using the DM flux estimation from Mikhail Kuznetsov given in %the 
Ref.\citep{Alcantara:2019sco} are 
shown with the gray solid line.

We estimate that the value of the GWs spectral break, which can be computed from the SHDM particle mass, could represent the  
smoking gun in the indirect searches for the SHDM with a certain mass. It is exciting that the GWs spectral break, $f_{\ast}$, for the range of the examined SHDM masses would be precisely within the sensitivity ranges of mid-band detectors such as BBO and LISA, see Fig.~\ref{fig:turnPointFreq_Mdm_constraints}.

In Fig.~\ref{fig:NANOGrav_comp} we map the extracted characteristic GW strain amplitude, $A$, by scanning the spectral index of the pulsar timing-residual cross-power spectrum, $\gamma_{N}$, using  
Eq.~\eqref{eq:GW_AmplEquation} on the recent finding of a stochastic common-spectrum process by NANOGrav. No matter what prior assumption we made on the initial values of SHDM mass, $G\mu$, $\alpha$ all obtained solutions  
intersect at the same value of the spectral index, $\gamma_{N}\simeq 4$. The slope of the lines in Fig.~\ref{fig:NANOGrav_comp} 
is driven by the choice of $M_{X}$ and $G\mu$. For $G\mu$ we took the obtained 
1$\sigma$ and 2$\sigma$ $G\mu$ from the fits to NANOGrav data \citep{Blasi:2020mfx}.

%%%%%%%%%%%%%%%%%%%%%%%%%%%%%%%%%%%%%
%%%%%%%%%%%%%%%%%%%%%%%%%%%%%%%%%%%%% 
\section{Summary and Conclusion}	
\label{sec:Conclusion}	

The abundance of SHDM can easily be dominant in the Universe today, with an SHDM density $\Omega_{\rm{shdm}}\sim\Omega_{\rm{dm}}$. This effect can be obtained by gravitational production that resembles the production of density fluctuations during inflation. The gravitational production of particles during inflation \cite{Chung:2001cb, Kannike:2016jfs} is the only experimentally verified DM production mechanism as the observed CMB fluctuations have exactly the same origin. 
This is because the production of SHDM during inflation gives rise to isocurvature perturbations that become sources of gravitational potential energy contributing to the tensor power spectrum of the CMB \citep{Chung:2004nh}. This implies a detectable primordial tensor-to-scalar ratio $r$ in the CMB power spectrum. 
At the end of inflation, a fraction of fluctuations are not stretched beyond the horizon but remain as particles because the inflation slows down. 
The weakness of gravitational interaction naturally explains the tiny initial abundance of WIMPzillas \cite{Kolb:1998ki}. Indeed, for such an abundance to be cosmologically relevant today, the X-particles must be supermassive. The combined [\citealt[][Planck satellite, together with]{Planck:2015sxf} \citealt[BICEP2 and the Keck array]{BICEP2:2015xme}] 95\% C.L. upper bound, $r \textless 0.07$, already constrains the X-particle mass to be $M_{X} \lesssim 10^{17}$ GeV in the limit of instantaneous reheating \citep{Garny:2015sjg}. For slightly less efficient reheating, this upper limit strengthens to $M_{X} \lesssim 10^{16}$ GeV.

There are several sources of constraints for the SHDM parameters. In the energy range of interest the mass, $M_{X}$, and  lifetime, $\tau_{X}$, are constrained by cosmic ray observations. 
The mass is subjected to cosmological constraints \citep{Kuzmin:1998kk,Chung:1998zb, Kolb:1998ki, Kuzmin:1999zk, Chung:2004nh, Gorbunov:2012ij} and GWs observations \citep{Samanta:2021mdm}.
The lifetime of the DM particles can be effectively constrained with the observed fluxes of various high-energy particles or with the upper limit on these fluxes. The upper limits to the gamma-ray flux obtained by Auger \citep{Kalashev:2016cre} and the non detection of events above $10^{11.3}$ GeV by Auger impose tight constraints \citep{Alcantara:2019sco} to the flux corresponding to this hypothetical SHDM component, see dark red bold line on Fig.~\ref{fig:DM_constraints}. 
The strongest limits for DM masses smaller than $\sim 10^{9}$ GeV are obtained from KASCADE-Grande \cite{Chianese:2021jke}, see red bold line on Fig.~\ref{fig:DM_constraints}. In \citep{Kalashev:2008dh} the constraints, see thin magenta solid line on Fig.~\ref{fig:DM_constraints}, have been put using the shape of charged cosmic-ray spectra. However, with the modern cosmic ray data this method is not as effective in constraining $\tau_{X}$ as neutrino \citep{Kuznetsov:2016fjt,Cohen:2016uyg} and gamma-ray \cite{Cohen:2016uyg,Murase:2012xs,Aloisio:2015lva,Esmaili:2015xpa,Kalashev:2016cre} data. Some fingerprints of the SHDM on GWs signal was discussed in \citep{Samanta:2021mdm,Bian:2021vmi}.

In the present paper, we have investigated the hypothesis of superheavy dark matter (SHDM) as a source of a subdominant component in the observed UHECR flux. The SHDM hypothesis is a viable candidate for DM in the Universe. 
Due to the expected small flux from the decay of SHDM particles the studies in this paper have been done in the context of the next generation of UHECR observatories, POEMMA, JEM-EUSO, and ZAP\citep{Romero-Wolf:2021wbw}, which are planned to have larger exposures compared with current ones. 
The limits obtained assuming the NFW DM profiles are weaker than those assuming the Einasto profiles. Given reports made in \citep{Chianese:2021jke} this makes NFW limits the loosest among commonly used DM distributions.

Taking into account all currently available constraints in the literature on the SHDM, one may conclude that if the DM flux is around $10^{-5}$ km$^{-2}$sr$^{-1}$y$^{-1}$ for masses in the range $10^{15} - 10^{17}$ GeV independently on the decay channel such SHDM hypothesis can be excluded. 
However, if there is some mechanism that may increase that flux by at least one order of magnitude, then there remains a window of opportunities to find these particles on the future POEMMA, JEM-EUSO and ZAP experiments.

Since the examined DM mass range significantly exceeds the sensitivity regions of the traditional DM
detection experiment, in view of the recent proposals \cite{Bian:2021vmi,Samanta:2021mdm} to search for the SHDM via GWs astronomy we put bounds on the possible DM signal probes with such detection technique.

%%%%%%%%%%%%%%%%%%%%%%%%%%%%%%%%%%%%%
%%%%%%%%%%%%%%%%%%%%%%%%%%%%%%%%%%%%% 
\begin{acknowledgments}
We thank Luis Anchordoqui and Olivier Deligny for useful comments regarding the SHDM flux and constraints estimation. 
MLeD acknowledges the financial support by the Lanzhou University starting
fund, the Fundamental Research Funds for the Central Universities
(Grant No. lzujbky-2019-25), National Science Foundation of China (grant No. 12047501) and the 111 Project under Grant No. B20063. 
\end{acknowledgments}
%---------------------------------------------

%%%%%%%%%%%%%%%%%%%%%%%%%%%%%%%%%%%%%
%%%%%%%%%%%%%%%%%%%%%%%%%%%%%%%%%%%%% 
%\section{appendix}

%---------------------------------------------
%\section{references}
%---------------------------------------------
%Produces the bibliography via BibTeX.

%\end{appendix}

\bibliography{DarkMatter_bibliography1}

\end{document}